# Focusing inversion technique applied to radar tomographic data

G. VIGNOLI[1] AND L. ZANZI[2]

[1] Università di Ferrara, Dipartimento di Scienze della Terra, Via Saragat 1, 4100 Ferrara, Italy
[2] Politecnico di Milano, Dipartimento di Ingegneria Strutturale, Piazza Leonardo da Vinci 32, 20133 Milano, Italy

**Summary**

Traveltime tomography is a very effective tool to reconstruct acoustic, seismic or electromagnetic wave speed distribution. To infer the velocity image of the medium from the measurements of first arrivals is a typical example of ill-posed problem. In the framework of Tikhonov regularization theory, in order to replace an ill-posed problem by a well-posed one and to get a unique and stable solution, a stabilizing functional (stabilizer) has to be introduced. The stabilizer selects the desired solution from a class of solutions with a specific physical and/or geometrical property; e.g., the existence of sharp boundaries separating media with different petrophysical parameters. Usually stabilizers based on maximum smoothness criteria are used during the inversion process; in these cases the solutions provide smooth images which, in many situations, do not describe the examined objects properly. Recently a new algorithm of direct minimization of the Tikhonov parametric functional with *minimum support* stabilizer has been introduced; it produces clear and focused images of targets with sharp boundaries. In this research we apply this new technique to real radar tomographic data and we compare the obtained result with the solution generated by the more traditional *minimum norm* stabilizer.

**Introduction**

Traveltime tomography is a very widely used technique: its uses go from mining exploration and petroleum reservoir characterization [1, 2] to non-destructive evaluation in engineering [3, 4]. To determine velocity distribution from measurements of traveltime is a non-unique problem and its numerical solution is unstable: small variations in the data can cause large variations in the solution. Commonly used inversion methods provide unique and stable solutions by introducing the appropriate stabilizing functional (stabilizer). The main aim of the stabilizer is to incorporate a priori knowledge in the inversion process. Over the last decade several different stabilizers have been introduced [5, 6, 7, 8, 9]. These new stabilizers permit reconstruction of blocky structures with abrupt change of properties. They generate clearer and more focused images of the anomalies than the conventional maximum smoothness functionals. For example, it was shown that the minimum support (MS) functional can be very useful in the solution of different geophysical inverse problems [8, 10, 11]. This particular functional selects the desired stable solution with the following characteristic: anomalies have sharp boundaries. The practical problem of focusing inversion with the MS stabilizer is that this functional is not quadratic, which complicates the minimization of the Tikhonov parametric functional. This difficulty has been faced using different approaches [8, 10, 12, 13, 14]. In this research we use a slightly different version of the algorithm developed by Zhdanov *et al* [11, 12, 14]; we show the result obtained applying this novel technique to real radar tomographic traveltimes. Moreover we compare MS solution with the result generated by the more traditional minimum norm (MN) stabilizer.

**MS stabilizing functional and weighting matrix**

Let us assume that the high-frequency limit is acceptable (i.e. the ray theory can be used instead of wave theory); let us suppose that the actual slowness distribution $s(\vec{r}) = 1/c(\vec{r})$, with $c(\vec{r})$ local velocity, is just a small perturbation of the background $s_b(\vec{r})$: $s(\vec{r}) = s_b(\vec{r}) + \Delta s(\vec{r})$. In this situation, as a first approximation, the raypath depends only on the background and the problem is linear. If we discretize the ground with a grid with constant slowness in the L cells, we can introduce a vector $\vec{m}$ of the model parameter as a vector of slowness within each cell: $\vec{m} = (s_1 s_2 ... s_L)$. The data vector $\vec{d}$ is formed by traveltimes: $\vec{d} = (\tau_1 \tau_2 ... \tau_N)$, where $\tau_i$ is the first arrival of the i-th ray. If we indicate the distance that the i-th ray travels within the j-th cell with $A_{ij}$, the expression for $\tau_i$ is:

$$\tau_i = A_{i1} s_1 + A_{i2} s_2 + ... = \sum_{j=1}^{L} A_{ij} s_j, \quad i = 1, 2, ... N. \quad (1)$$

Using matrix notation, we can write: $\vec{d} = \hat{A}\vec{m}$.

A common way to solve inverse problems is by minimization of the Tikhonov parametric functional:

$$P^\alpha(m) = \phi(m) + \alpha s(m) \quad (2)$$

that combines least square data misfit functional: $\phi(m) = \|A(m) - d\|_{L_2}^2$ and the stabilizer $s(m)$, whose function is to select a correctness subset $M_c$ from the space of all possible models $M$. There are several different choices for the stabilizer and of course different stabilizers produce different solutions. In this research, we analyze:

- the MN stabilizer ($s_{MN}$) which is proportional to the difference between the model $m$ and an appropriate a priori model $m_{apr}$:

$$s_{MN}(m) = \|m - m_{apr}\|_{L_2}^2; \quad (3)$$

- the MS stabilizer ($s_{MS}$) which is equal to the area (support) where the difference between the current model $m$ and the a priori model $m_{apr}$ is non-zero:

$$s_{MS}(m) = \int_V \frac{(m - m_{apr})^2}{(m - m_{apr})^2 + e^2} dv, \quad (4)$$

where e is the focusing parameter. It is easy to demonstrate that $s_{MS} \to \text{support}(m - m_{apr})$ if $e \to 0$ [8].

Before selecting the correctness subset $M_c$, one can act on $M$ with a linear transformation. So we look for the stable solution in the space $\{\vec{m}^w\}$ with $\vec{m}^w = \hat{W}_m \vec{m}$. We select $\hat{W}_m$ as the square root of the sensitivity matrix:

$$\hat{W}_m = \text{diag}(\hat{A}^T \hat{A})^{1/2} = \text{diag}\left(\sqrt{\sum_i (A_{ij})^2}\right). \quad (5)$$

As a result, we obtain a uniform sensitivity of the traveltimes to different model parameters [10]. If we take into account that $\sum_i A_{ij}$ is the total distance that all rays travel in the cell j-th, it is clear that these weights make all cells equally "important" and it doesn't matter if a cell is crossed by many or few rays. If we do not use any weighting matrix, data are more sensitive to the velocity value of those cells crossed by many rays.

The regularization parameter α in (2) describes a trade-off between the best fitting and the most reasonable stabilization. How do we select the optimal α? We choose a set of values $\{\alpha_k\}$, then we find for each $\alpha_k$ the corresponding model $m_{\alpha_k}$ minimizing $P^{\alpha_k}(m)$ and finally we calculate the misfit $\phi(m_{\alpha_k})$. The optimal value is the number $\alpha_{k_0}$ such that we have

$$\phi(m_{\alpha_{k_0}}) = \delta, \tag{6}$$

with $\delta$ noise level in the observed data.

Often we know a priori that the model parameters are bounded: $\vec{m}^- \leq \vec{m} \leq \vec{m}^+$, where $\vec{m}^-$ and $\vec{m}^+$ are the lower and upper bounds. In our algorithm, we try to force the inversion to produce a model with this characteristic: at the *beginning* of each iteration, we reset the velocity values in order to have $\vec{m}_i \geq (\vec{m}^-)_i = \frac{1}{30}\frac{\text{ns}}{\text{cm}}$.

**Radar traveltimes inversion**

We compare inversion results obtained using the two stabilizers discussed overleaf. Data were radar first arrivals collected over a brick wall with two cavities, whose position and dimension (11 cm x 14.4 cm) were known. The antenna frequency was 1.6 GHz. The receiver was located on one side of the wall in 49 positions spaced 2.5 cm apart. For every receiving antenna position the transmitter was moved through 49 positions along a vertical profile on the other side of the wall; thus the available traveltimes were ~49x49=2401. The antenna positions were monitored by a new prototype for automatic positioning. An interactive program for semiautomatic picking, developed at the Politecnico di Milano, was used to extract the traveltimes.

We choose to terminate the minimization process of $P^\alpha(m)$ when the misfit condition (6) is reached, assuming that the noise level in the observed data is 3.5%.

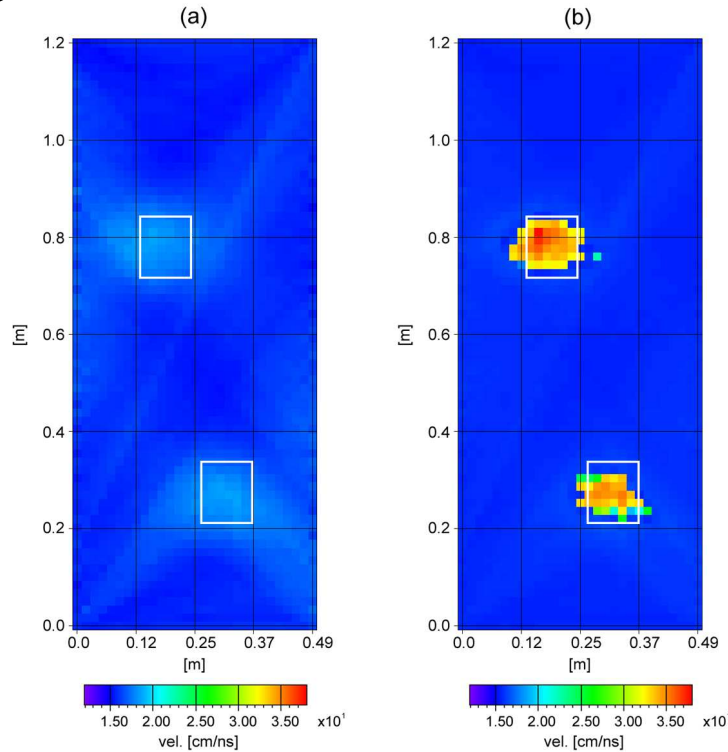

Figure 1: a comparison of minimum norm (a) and minimum support (b) inversion results.

Figure 1a shows MN result: it has the typical relatively smooth model parameter distribution, the image is diffused and unfocused, the position of the two cavities (indicated by the two white frames) is not correctly detected.

In the MS solution (figure 1b), cavity positions are quite well determined. Blurring effects, very common in this kind of tomography, are almost absent. Dimensions are lightly underestimated: in the case of the void on the right, probably because of the low ray coverage in that area. Moreover, in general, we should take into account that the velocity within the cavities is twice the background velocity, thus the actual slowness is not really a small perturbation of the background as we assume.

**Conclusions**

Stabilization of inversion methods with smoothing functionals can result in oversmoothed reconstruction of object properties. We suggest the use of the $s_{MN}$ to preserve sharp features in inverted models. The example shows that this approach can recover sharp contrast velocity distributions when prior information about the anomalies is available. A more sophisticated way to insure that model parameters belong to some interval (e.g. introducing a logaritmic parametrization) and to include a ray-tracing tool would make the result even better.